\newcommand\aj{Astronomical Journal}%
\newcommand\araa{Annual Review of Astron \& Astrophysics}%
\newcommand\apj{Astrophysical Journal}%
\newcommand\apjl{Astrophysical Journal Letters}%
\newcommand\apjs{Astrophysical Journal Supplement}%
\newcommand\aap{Astronomy \& Astrophysics}%
\newcommand\aaps{Astronomy \& Astrophysics Supplement}%
\newcommand\mnras{Monthly Notices of the Royal Astronomical Society}%
\documentclass[
    ,final            
  ]
  {aipproc}

\layoutstyle{6x9}

\begin{document}

\title{SN~1996cr: Confirmation of a Luminous Type~IIn Supernova in the Circinus Galaxy}

\classification{95.85.-e, 95.85.Bh, 95.85.Kr, 95.85.Nv, 95.85.Pw, 97.10.Fy, 97.60.Bw}

\keywords      {SNe, SN1996cr, X-ray, Radio, Optical, Gamma-Ray}

\author{F.~E.~Bauer}{
  address={{\it Chandra} Fellow, Columbia Astrophysics Laboratory, 550 W. 120th St., Columbia University, New York, NY 10027}
}

\author{S.~Smartt}{
  address={Department of Physics and Astronomy, Queen's University Belfast, Belfast, BT7 1NN Northern Ireland, UK}
}

\author{S.~Immler}{
  address={Goddard Space Flight Center, Code 662, Greenbelt, MD 20771, USA}
}

\author{W.~N.~Brandt}{
  address={Department of Astronomy \& Astrophysics, 525 Davey Lab, The Pennsylvania State University, University Park, PA 16802.}
}

\author{K.~W.~Weiler}{
  address={US Naval Research Lab, 4555 Overlook Ave., SW, Washington, DC 20375}
}

\begin{abstract}
We have recently confirmed SN~1996cr as a late-time type~IIn supernova
(SN) via VLT spectroscopy and isolated its explosion date to
$\sim$1~yr using archival optical imaging. We briefly touch upon here
the wealth of optical, \hbox{X-ray}, and radio archival observations
available for this enigmatic source. Due to its relative proximity
(3.8$\pm0.6$~Mpc), SN~1996cr ranks among the brightest \hbox{X-ray}
and radio SNe ever detected and, as such, may offer powerful insights
into the structure and composition of type~IIn SNe. We also find that
SN~1996cr is matched to GRB 4B~960202 at a 2--3$\sigma$ confidence
level, making it perhaps the third GRB to be significantly associated
with a type~II SN. We speculate on whether SN~1996cr could be an
off-axis or ``failed'' GRB.
\end{abstract}

\maketitle


\vspace{-0.35in}
\section{Background}
\vspace{-0.05in}
SN~1996cr was discovered by {\it Chandra} as Circinus Galaxy X-2
\citep{Bauer2001}, an ultraluminous \hbox{X-ray} source which exhibited many
traits of a young supernova (SN) interacting with dense circumstellar
material (CSM).
The Circinus Galaxy (Circinus) is a nearby (3.8$\pm$0.6 Mpc), massive
spiral galaxy situated close to the Galactic plane
($b=3$\mbox{$.\!\!^\circ$}8). Due to its location in the sky,
\hbox{SN~1996cr} suffers from $N_{\rm H}\sim3\times10^{21}$~cm$^{-2}$
($A_{\rm V}\sim1.5$) due to our own Galaxy as determined by radio and
infrared measurements, and $N_{\rm
H}\sim(3$--$5)\times10^{21}$~cm$^{-2}$ ($A_{\rm V}\sim$1.5--2.5)
internally based on \hbox{X-ray} column density constraints.

\vspace{-0.15in}
\section{Confirmation and Explosion Date Constraints}
\begin{figure}
  \hglue-0.6cm{\includegraphics[height=.25\textheight]{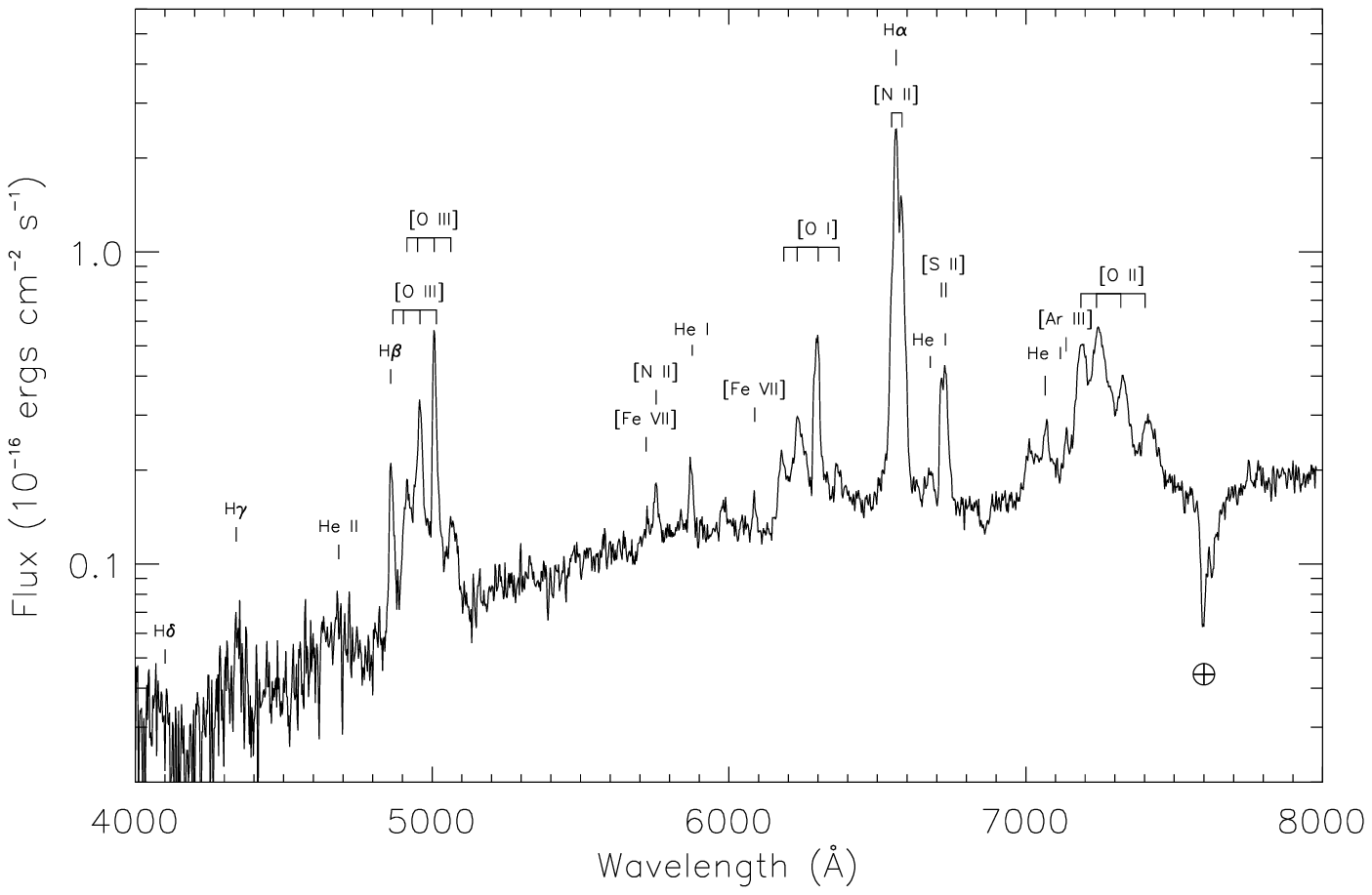}}\hglue0.1cm{\includegraphics[height=.24\textheight]{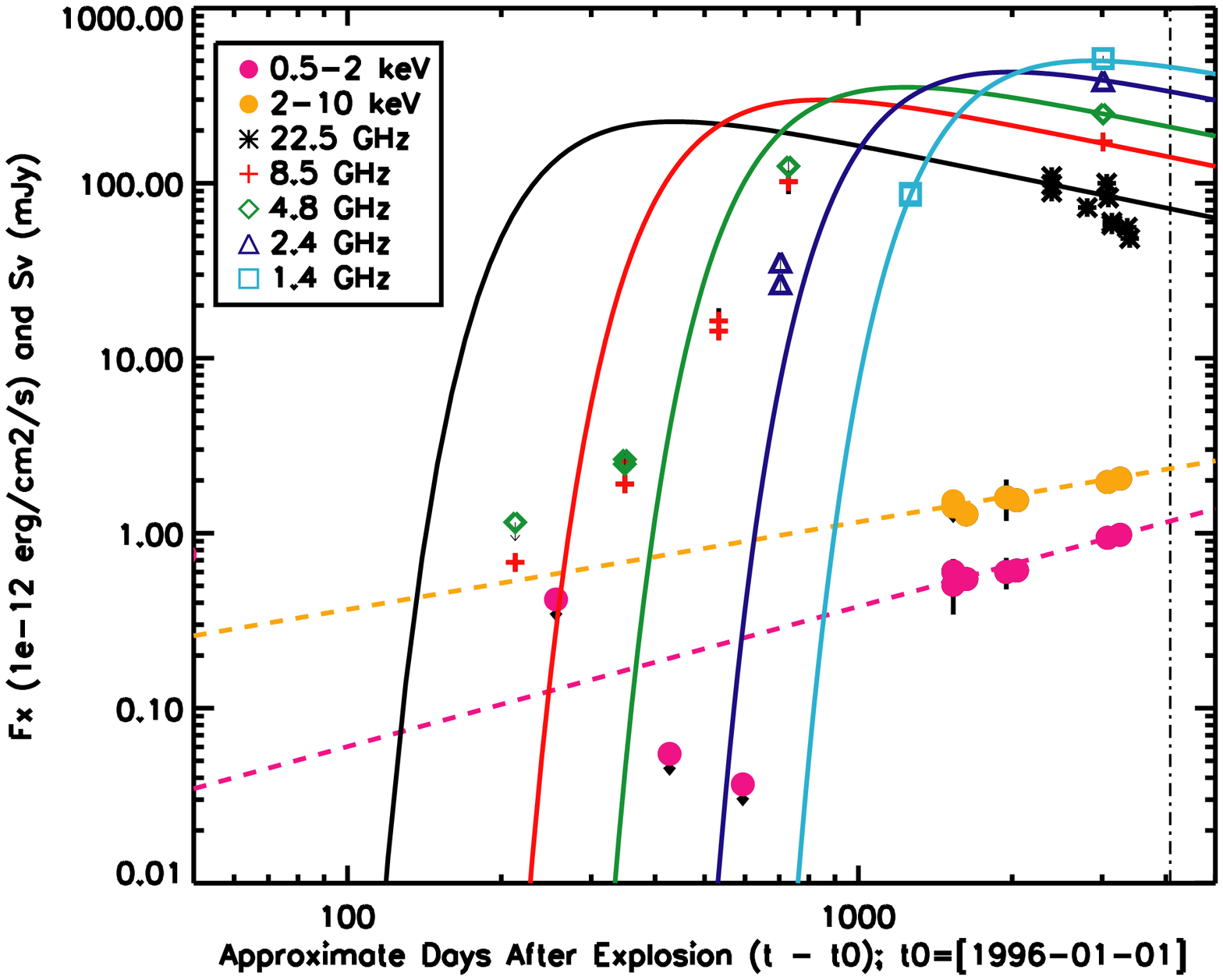}}
  \caption{{\it Left:} VLT FORS1 spectrum of \hbox{SN~1996cr} taken on
  2006 Jan 26. The spectrum features broad, asymmetric emission
  complexes, as well as a strong narrow H$\alpha$ emission line. Such
  features are typical of late-time SNe interacting with dense
  circumstellar material (i.e., type~IIn).  {\it Right:} Light curves
  for both the radio (above) and \hbox{X-ray} (below) data. The radio
  data were fit with a relatively standard ejecta-CSM interaction
  model, which does well with the late-time radio points, but has
  difficulty fitting the unusual early-time radio points. Due to the
  lack of late-time points, the slope of decline is not well
  determined (aside from the somewhat erratic 22~GHz data, it is not
  clear whether the radio emission has even ``rolled over'' yet).
  Note that even forcing a rather extreme early-time Synchrotron-Self
  Absorption model on \hbox{SN~1996cr} is not able to acceptably fit
  the rather odd early data points.  The \hbox{X-ray} data from {\it
  Chandra} and {\it XMM-Newton} show a relatively strong and
  significant rise between 2000-2004 both in the soft and hard bands;
  this is very atypical for \hbox{X-ray} detected SNe. The strong {\it
  ROSAT} soft-band constraints and unusual early-time radio data hint
  that the progenitor of \hbox{SN~1996cr} may have blown out a
  relatively low-density cavity immediately prior to its demise.}
\label{fig}
\end{figure}

\vspace{-0.05in}
A high quality VLT FORS~I spectrum of SN~1996cr taken on 2006 Jan 26
(Fig.~\ref{fig}) confirms it as a type~IIn SN. The spectrum is
dominated by several narrow emission lines (e.g., H$\alpha$, [N\,{\sc
ii}], and [S\,{\sc ii}]; FWHM$\sim$ 700~km~s$^{-1}$), as well as
several strong, complex emission lines of [O\,{\sc i}], [O\,{\sc ii}],
[O\,{\sc iii}] comprised of several partially blended broad emission
lines (FWHM$\sim$2000--3000~km~s$^{-1}$) which are red- and
blue-shifted with respect to rest wavelengths. Such features imply
that the rapidly moving SN material is plowing into a clumpy dense CSM
and driving slower shocks into them. The strong, distinct Oxygen peaks
suggest the ejecta are perhaps located in a few asymmetric,
Oxygen-rich shells or rings.

Narrow-band imaging observations of Circinus with the Taurus
Fabry-Perot instrument on the Anglo Australian Telescope between 1995
February 21--28 and 1996 March 15--20 constrain \hbox{SN~1996cr}'s
explosion date to within $\sim$1~yr. From these images, we estimate
\hbox{SN~1996cr} to be $M_{\rm V}\approx-10.2$, or at least $M_{\rm
Vc}\approx-15$ when corrected for expected extinction. This provides a
strong lower limit, given the explosion date and extinction
uncertainties, and indicates that
\hbox{SN~1996cr} was at least of average peak brightness among type~II
SNe \citep{Patat1994}. While the optical data provide the tightest
constraints on the explosion date, the plentiful archival \hbox{X-ray} and
radio data offer invaluable constraints on the surrounding
environment. We plot together the soft and hard \hbox{X-ray} fluxes, as well
as radio flux densities in a variety of bands (Fig.~\ref{fig}). The
early \hbox{X-ray} and radio upper limits imply the presence of either strong
early absorption or a low-density cavity immediately surrounding the
progenitor. When finally detected, the 0.5--2 and 2--10 keV \hbox{X-ray}
fluxes are best-fitted as $\propto$ $t^{1.0}$ and $t^{0.7}$,
respectively, whereas nearly all SNe are expected to decline as
$\propto$ $t^{-1}$--$t^{-0.4}$
\citep{Chevalier1994} (although observationally there is more scatter
\citep{Immler2005}). This decade-long rise in the \hbox{X-ray} is most unusual 
and has only been observed for \hbox{SN~1987A}
\citep{Park2006} and marginally for \hbox{SN~1978K}
\citep{Schlegel2004}; thus \hbox{SN~1996cr} may be an intermediate
object between the extremes of \hbox{SN~1987A} and more typical,
luminous SNe. \hbox{SN~1996cr}'s rise at radio wavelengths exhibits an
unusual convex spectral shape which is poorly fit by conventional
models \citep{Weiler2002} and argues against prolonged early
absorption. The strong \hbox{X-ray} and radio emission imply that the CSM is
quite dense ($>10^{-4}$~$M_\odot$~yr$^{-1}$), while the lack of broad
H$\alpha$ suggests that the Hydrogen shell was likely cast off prior
to explosion. This points to a massive, stripped-core progenitor,
perhaps similar to the luminous blue variable $\eta$ {\it
Carinae} \citep{Gal-Yam2007}. Notably, the most recent \hbox{X-ray} and radio
data demonstrate that \hbox{SN~1996cr} is already one of the brightest
\hbox{X-ray} and radio SN on the sky and will likely climb even higher.
Thus future monitoring observations should place important
constraints on the evolution and nature of the progenitor.

\vspace{-0.15in}
\section{\hbox{SN~1996cr} as a Gamma-Ray Burst (GRB)?}
\vspace{-0.05in}
Within temporal and positional errors we find that \hbox{SN~1996cr}
coincides with BATSE/GRB 4B~960202 \citep{Meegan1996}. The strict
prescription of \citet{Wang1998} yields a probability of 4.6\%,
although this probability is dominated by a few very poorly located
sources, while our candidate has a smaller than average $1\sigma$
positional error of 0\mbox{$.\!\!^\circ$}88; the probability that one
of the 65 possible candidates instead falls within
$3\times$0\mbox{$.\!\!^\circ$}88 of
\hbox{SN~1996cr} is only 0.3\%. While intriguing, this association
remains weak due to the wide range in explosion dates. The link is
plausible, however, since the potential progenitor of
\hbox{SN~1996cr} appears stripped of its outer envelope, a necessary
condition postulated for GRB progenitors \citep{Woosley2006}. At the
distance of Circinus, 4B~960202 would have an observed luminosity of
$7\times10^{45}$~ergs~s$^{-1}$, making it the least luminous GRB
detected to date; typical GRBs have $\sim10^{51}$~ergs~s$^{-1}$ and
even the meager GRB~980425 was $8\times10^{47}$~ergs~s$^{-1}$
\citep{Galama1999}. 4B~960202's apparently weak luminosity could have
several causes. For instance, it could have been intrinsically weak;
the underlying physical model that produces GRBs is believed to
generate a broad range of luminosities. 4B~960202 could simply lie on
the weak tail, perhaps even in the regime of ``failed'' GRBs, whereby
the progenitor may not have shed enough of its outer envelope, forcing
the jet to expend the bulk of its energy tunneling out. Or we could
simply be viewing the GRB off-axis and only be seeing a very small
percentage of the overall beamed energy. We note that there are two
other type~II SNe found to be marginally coincident with GRBs
\citep{Rigon2003}: SN~1999E (GRB~980910, $z=0.0258$) and SN~1997cy
(GRB~970514, $z=0.059$). While the progenitors of these two SN were
likely massive stars, both SNe show broad H$\alpha$ indicative of a
relatively intact progenitor which is at odds theoretically
with GRB formation mechanisms \citep{Woosley2006}. Both GRBs are
also significantly underluminous. Taken together, these objects may
signal that the physical engines which inevitably drive GRBs may
be relatively common in core-collapse SNe.


\bibliographystyle{aipproc}



\IfFileExists{\jobname.bbl}{}
 {\typeout{}
  \typeout{******************************************}
  \typeout{** Please run "bibtex \jobname" to optain}
  \typeout{** the bibliography and then re-run LaTeX}
  \typeout{** twice to fix the references!}
  \typeout{******************************************}
  \typeout{}
 }

\end{document}